\documentclass[hidelinks, conference,letterpaper]{IEEEtran}

\usepackage{latexsym}
\usepackage{arydshln}
\usepackage{cite}
\usepackage{bm}
\usepackage{subfigure}
\usepackage{amsmath}
\usepackage{extarrows}
\usepackage{amssymb}
\usepackage{multirow}
\usepackage{graphicx}
\usepackage{color}
\usepackage{enumerate}
\usepackage{bookmark}
\graphicspath{{figure}}

\usepackage[utf8]{inputenc}
\usepackage[T1]{fontenc}
\usepackage{url}
\usepackage{ifthen}
\usepackage{stfloats}

\newcommand{\mr}{\mathrm}

\newcommand{\BE}{\begin{equation}}
\newcommand{\EE}{\end{equation}}
\newcommand{\BS}{\begin{subequations}}
\newcommand{\ES}{\end{subequations}}
\renewcommand{\bf}{\bm}

\ifCLASSINFOpdf
\graphicspath{{figure/}}
\else
\fi
\begin{document}

\title{Irregularly Clipped Sparse Regression Codes}

         \author{%
  \IEEEauthorblockN{ Wencong~Li}
  \IEEEauthorblockA{
                    Japan Advanced Institute of\\ Science and Technology\\
                    923-1292 Nomi, Japan\\
                    Email:  liwencong@jaist.ac.jp}
  \and
  \IEEEauthorblockN{Lei~Liu}
  \IEEEauthorblockA{Japan Advanced Institute of\\                   Science and Technology\\
                    923-1292 Nomi, Japan\\
                    Email: leiliu@jaist.ac.jp}
\and
  \IEEEauthorblockN{Brian~M.~Kurkoski}
\IEEEauthorblockA{Japan Advanced Institute of\\                       Science and Technology\\
                    923-1292 Nomi, Japan\\
                    Email: kurkoski@jaist.ac.jp}
}
\maketitle

\begin{abstract}
Recently, it was found that clipping can significantly improve the section error rate (SER) performance of sparse regression (SR) codes if an optimal clipping threshold is chosen.
In this paper, we propose irregularly clipped SR codes, where multiple clipping thresholds are applied to symbols according to a distribution, to further improve the SER performance of SR codes. Orthogonal approximate message passing (OAMP) algorithm is used for decoding. Using state evolution, the distribution of irregular clipping thresholds is optimized to minimize the SER of OAMP decoding. As a result, optimized irregularly clipped SR codes achieve a better tradeoff between clipping distortion and noise distortion than regularly clipped SR codes. Numerical results demonstrate that irregularly clipped SR codes achieve 0.4 dB gain in signal-to-noise-ratio (SNR) over regularly clipped SR codes at code length$\,\approx2.5\!\times\! 10^4$ and SER$\,\approx10^{-5}$. We further show that irregularly clipped SR codes are robust over a wide range of code rates.
\end{abstract}



\section{Introduction}

Sparse regression (SR) codes were introduced and studied by Barron and Joseph \cite{Joseph2012} and were shown to achieve the capacity of the additive white Gaussian noise (AWGN) channel when the code length and the rate of non-zero symbols go to infinity \cite{Rush2017,Barbier2015, Barbier2017}. Furthermore, the code rate can be easily changed by adjusting the matrix size directly rather than rebuilding the parity check matrix carefully as in low density parity check (LDPC) codes \cite{Gallager1962} because there is no sparse/tree-like requirement on the measurement matrix. The low-complexity fast Fourier transform (FFT) algorithm can be used to reduce the complexity and efficient approximate message passing (AMP) type algorithms can be used for decoding to guarantee a Bayes-optimal solution, i.e., the minimum mean square error (MMSE) solution \cite{Rush2017,Barbier2015,Barbier2017}. However, SR codes have a weakness at finite code length. When the code length and the compression rate are finite, the performance of SR codes is far from the capacity, and much worse than the widely used LDPC codes.

 To improve the performance and competitiveness of SR codes, several techniques have been introduced. Spatial-coupling, originally introduced for LDPC codes, was used in SR codes \cite{Barbier2017} to obtain a Bayes-optimal decoding performance for any compression rate. Recently, it was found that regular clipping, a technique that is generally used to reduce the peak-to-average-power ratio (PAPR), can also improve the section error rate (SER) performance of SR codes significantly \cite{Liang_cl2017}. When a signal is clipped, we compensate by scaling the clipped signal to maintain a fixed transmit power; the intuition of clipped SR codes is that this scaling is more beneficial than the penalty of distortion due to clipping. 
Therefore, there exists a tradeoff between the clipping distortion and noise distortion. An optimal clipping threshold can bring a significant gain in SER performance for SR codes at finite code length, sometimes achieving 4 dB in SNR at code length $10^4$.  
However the performance is still not very close to the capacity (about 2 dB at code length $10^4$).  

This paper considers irregular clipping of SR codes. In addition, a permuted discrete cosine  transform (DCT) matrix is used for the SR code measurement matrix, rather than an i.i.d. Gaussian matrix as in previous work. Thus, low-complexity fast DCT and inverse DCT (IDCT) algorithms can be applied to reduce the encoding and decoding complexity of SR codes. Orthogonal AMP (OAMP) algorithm \cite{Ma2016} (see also closely-related earlier works in \cite{MaSPL2015a, Ma_SPL2015b}) is exploited for the decoding of the clipped SR codes.  It was shown in \cite{Rangan2016, Takeuchi2017} that state evolution (SE) can asymptotically characterize the performance of OAMP algorithm, similar to extrinsic information transfer (EXIT) chart in iterative decoding \cite{Brink2001, Brink2004}. In this paper, the main contributions are summarized as follows.  
\begin{itemize}
    \item  To further improve the SER performance of regularly clipped SR codes, we propose an irregularly clipped SR code. In irregularly clipped SR codes, a new irregular clipping method applies different clipping thresholds to different symbols, which provides a larger optimization space and thus results in better SER performance.
    
    \item Based on the SE analysis for the OAMP decoding, the distribution of irregular clipping thresholds is optimized to minimize the SER of irregularly clipped SR codes, which is a concave optimization and thus can be solved by standard convex optimization tools.
    
    \item We provide numerical results to demonstrate the efficiency of irregularly clipped SR codes. Specifically, comparing with the existing regularly clipped SR codes, the proposed irregularly clipped SR codes achieve about 0.4 dB gain in signal-to-noise-ratio (SNR) at code length $\approx\! 2.5\!\times\! 10^4$ and SER  $\approx10^{-5}$. Furthermore, the irregularly clipped SR codes are robust in a wide range of code rates, e.g., from 0.2 (low rate) to 1 (high rate).  
\end{itemize}  

 The findings in this paper provide a promising direction to significantly enhance the performance of SR codes.

\section{ {Preliminaries}} 
 {In this section, we briefly introduce  SR codes, regularly  clipped SR codes and OAMP decoding.}

 {Consider an AWGN channel:
\BE\label{equ.recovery_model}
 \bm{y}=\bm{c}+\bm{n} \vspace{1.5mm},
\EE
where $\bf{c}\in \mathbb{R}^{M}$ is an encoded vector and $\bf{n}\sim \mathcal{N}(\bf{0},\sigma_n^2\bf{I}_{M})$ is a Gaussian noise vector. We define ${\rm SNR}=\sigma_n^{-2}$. The goal is to recover $\bm{c}$ via the received signal vector $\bm{y}$.}

\subsection{ {Sparse Regression (SR) Codes}}
 {An SR code is generated by \cite{Rush2017,Barbier2017}:
\BE\label{equ.linear_system}  
  \bm{c} = \bm{Ax}, 
\EE
where ${\bm A}\!\in\! \mathbb{R}^{M\!\times\! N}$ is a measurement matrix with compression rate $\delta\!=\!M/N\!\leq \!1$, and ${\bm x}$ is an information-carrying sparse vector that contains $L$ sections. Each section is of length $B$ and has a single non-zero entry with amplitude $\sqrt{B}$. An example of ${\bm x}$ is as follows:
\BE\label{Eqn:block_sparse}
\!\!\!\!\Phi\!: \bf{x}^{\mr{T}}\!\!=\![\overbrace{0,\dots,0,\!\sqrt{\!B}}^{\mr{section}\; 1}|\overbrace{\dots,0,\!\sqrt{\!B},0}^{\mr{section}\;2}|\dots,|\overbrace{\!\sqrt{\!B},0,\dots,0}^{\mr{section}\;L}].
\EE
The positions of the non-zero symbols in $\bf{x}$ carry the information. In this code, the code length is $M$ and the amount of information is $L{\rm log}B$. Thus, the code rate is given by $R={L{\rm log}B}/{M}$.}
 
 {For theoretical analysis in \cite{Rush2017,Barbier2017}, ${\bm A}\in \mathbb{R}^{M\times N}$ in \eqref{equ.linear_system}   is assumed to have independent identically distributed Gaussian (IIDG) entries over $\mathcal{N}(0, N^{-1})$. In this paper, ${\bm A}$ is generated by randomly selecting $M$ rows from the $N \times N$ DCT matrix. Hence, the low-complexity fast DCT algorithm can be used in encoding and decoding. In addition, OAMP \cite{Ma2016} can be applied to such non-IID sensing matrices. Furthermore, it is shown that the performance of partial DCT matrices is better than IID Gaussian matrices \cite{Ma_SPL2015b}. Apart from that, it was proved that the OAMP decoding is potential Bayes optimal and obtains the maximum \emph{a posteriori} (MAP) solution of \eqref{equ.recovery_model}. }

SR codes work well at a very large length but are not as good as expected in more practical scenarios (e.g., limited code length and high code rates). Therefore a new method to improve the performance is needed.


\subsection{ {Regularly Clipped SR Code }}\label{Sec:RC_SRC}
 Clipping is normally applied to reduce PAPR. 
 It sets the signals whose absolute value is larger than a given threshold to the threshold and then normalizes the transmitted signals to keep the same power.

 {Let ${\mr{clip}}(\cdot)$ be a symbol-by-symbol clipping given by 
\BE\label{Eqn:clip}
{\mr{clip}}_{\epsilon}\left( z \right) =\left\{ \begin{array}{l}
\epsilon, \qquad \;z>\epsilon\\
z, \qquad  |z| \leq \epsilon\\
-\epsilon, \quad\;\; z<-\epsilon
\end{array} \right..
\EE
The clipping ratio (CR) of ${\mr{clip}}(\cdot)$  in \eqref{Eqn:clip} is defined as ${\rm CR} = 10\rm{log}_{10}(\epsilon^2/E\{z^2\})$. Then, the codeword ${\bm c}$ is  given by a normalized clipping function defined as \cite{Liang_cl2017}:
\BE\label{Eqn:C_SRC}
 {\bm c}    = \alpha \, {\mr{clip}}_{\epsilon}(\bf{Ax}),
\EE
where $ \alpha= \frac{1}{\sqrt{{\rm E} \{\|{\mr{clip}}(\bf{Ax})\|^2\}/M}}$ is a normalizing factor. We call \eqref{Eqn:C_SRC} a regularly clipped SR code since all the elements in $\bf{Ax}$ are clipped with the same threshold $\epsilon$. Intuitively, $\alpha$ makes a power compensation for the clipping operation, which introduces a tradeoff between the clipping distortion and channel noise distortion.}

\begin{figure}[t]
    \includegraphics[width =8.5cm]{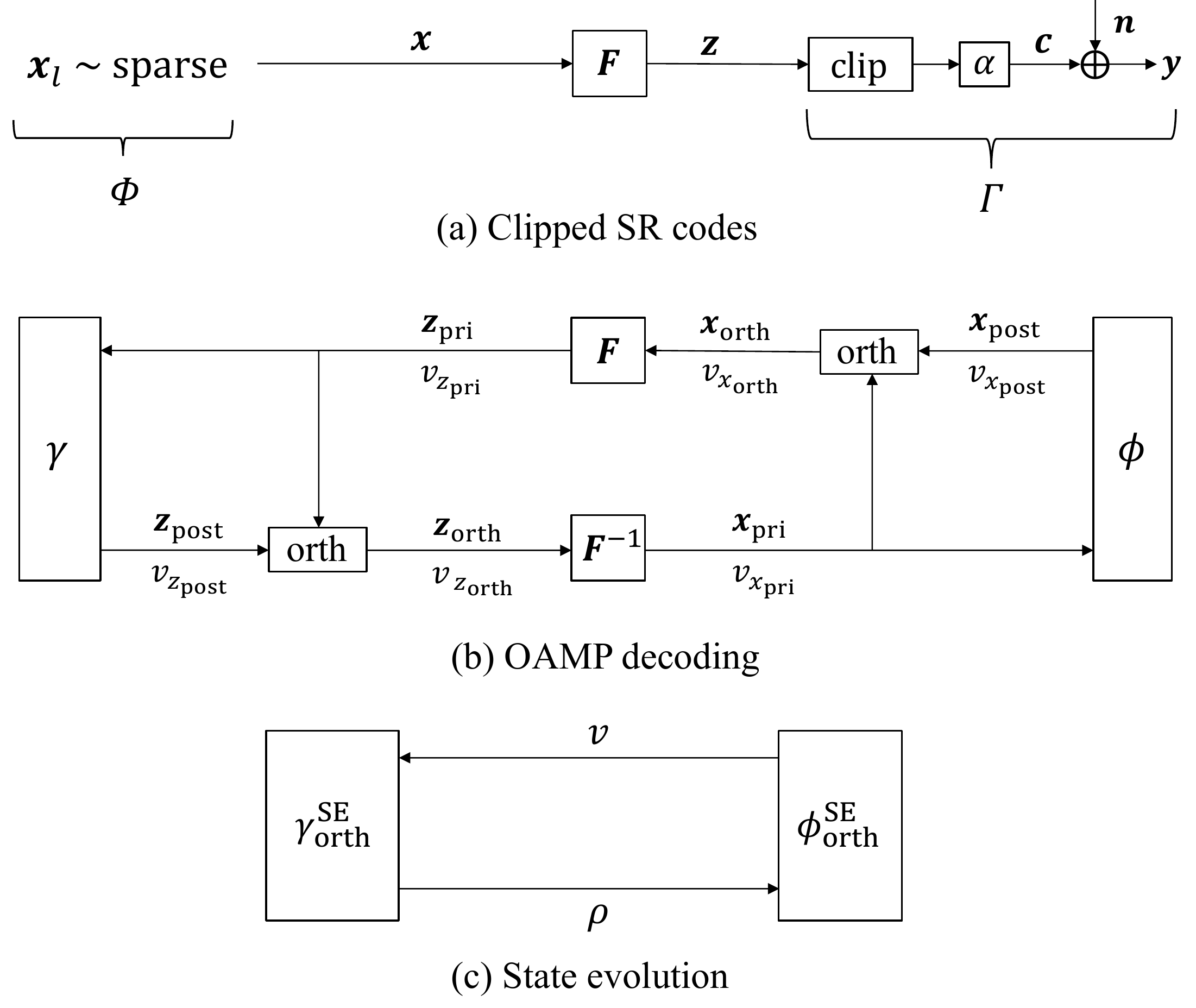}
    \caption{ {Graphical illustrations of (a) clipped SR codes and AWGN channel, (b) OAMP decoding and (c) state evolution of OAMP. Functions $\gamma$ and $\phi$ in (b) are respectively corresponding to $\Gamma$ and $\Phi$ in (a).  $\bf{F}$ and $\bf{F}^{-1}$ are DCT and inverse DCT (IDCT) corresponding to the transfer $\bf{F}$ in (a). Orthogonalization ``orth" is used to solve the correlation problem in iterative process.}}
    \label{Fig.SR_block}
\end{figure}

 {For simplicity, we rewrite the problem as:
\begin{subequations}\label{equ.clipped_SR}
\begin{align}
&\Gamma:\quad\bm{y}= \alpha \, {\rm clip}_{\epsilon}(\bm{z})+\bm{n},  \vspace{1.5mm}\label{equ.clipped_SR_a}\\ 
& \Psi:\quad\bm{z} = \bm{Fx}, \label{equ.clipped_SR_b}
\\
&\Phi:\quad
\bf{x}_l \sim P(\bm{x}_l),\;\; \forall l,  \label{equ.clipped_SR_c}
\end{align}
\ES
where $\Gamma$ contains a symbol-by-symbol clipping and an AWGN channel,  $\Psi$ is a DCT process ($\bm{F}$ is a DCT matrix), $\bm{x}_l$ denotes the $l$-th section in $\bf{x}$, and $\Phi$ the block sparse modulation in \eqref{Eqn:block_sparse}. Fig. \ref{Fig.SR_block}(a) gives a graphical illustration of the clipped SR codes.}

 {Optimized clipping can significantly improve the SER performance of SR codes \cite{Liang_cl2017}. For example,  as shown in Fig. \ref{Fig.simulation_result}, when code length $=10^{4}$ and $R=0.5$, the clipped SR code with optimal clipping threshold achieves 4dB gain in SNR at SER $=10^{-5}$ comparing the un-clipped case. However, it is still 2dB away from the Shannon limit  (0 dB for $R=0.5$) of the system in \eqref{equ.recovery_model}.

\subsection{ {OAMP Decoding}}
 {Fig. \ref{Fig.SR_block}(b) gives an illustration of the OAMP decoding process, which consists of the following modules:
\begin{itemize}
    \item De-clipping process $\gamma(\cdot)$ solves the clipping-and-AWGN constraint $\Gamma$ in \eqref{equ.clipped_SR_a}.
    \item De-modulation process $\phi(\cdot)$ for the sparse modulation constraint $\Phi$ in \eqref{equ.clipped_SR_c}.
    \item $\bf{F}$ and $\bf{F}^{-1}$ correspond to the transfer constraint in \eqref{equ.clipped_SR_a}.
    \item The orthogonalization ``orth" solves the correlation problem in  iterative process and ensures the correctness of state evolution \cite{Ma2016}.
\end{itemize}
Their expressions are given by 
\BS\label{equ.OAMP}
\begin{align}
\!\!\!\!&{\rm Declip:}&  \bm{z}_{\rm post}&=\gamma_\epsilon(\bm{z}_{\rm pri})={\rm E} \{\bm{z}|\Gamma,\bm{z}_{\rm pri},v_{z_{\rm pri}}\},\label{equ.OAMP.Declip}
\\
&& v_{\rm post}&=\! \gamma_\epsilon^{\rm SE}(v_{z_{\rm pri}})\!=\!{\rm var} \{ \bm{z}|\Gamma\!,\bm{z}_{\rm pri},v_{z_{\rm pri}}\},
\label{equ.OAMP.Declip_v} \\ \nonumber\\
 &  {\rm Orth:}& v_{z_{\rm orth}} &\!=\!{\cal O}_{\rm SE}\big(\delta v_{z_{\rm post}}\!\!+\!(1\!-\!\delta)v_{z_{\rm pri}},v_{z_{\rm pri}}\big),\\  
 & & \bm{z}_{\rm orth} &= {\cal O}(\bm{z}_{\rm post},v_{z_{\rm post}};\bm{z}_{\rm pri},v_{z_{\rm pri}}), \\ \nonumber
  \\
&{\rm IDCT:} &   \bm{x}_{\rm pri} &=  \bm{F}^{-1}\bm{z}_{\rm orth},\quad    v_{x_{\rm pri}}  = v_{z_{\rm orth}}, \label{equ.OAMP.IFFT}
\\\nonumber\\
 &{\rm Demod:} & \bm{x}_{\rm post} &= \phi(\bm{x}_{\rm pri})= {\rm E} \{ \bm{x}|\Phi,\bm{x}_{\rm pri},v_{x_{\rm pri}}\}, \\ 
 && v_{\rm post}& =\!\phi^{\rm SE}\!(v_{x_{\rm pri}}) \!=\!{\rm var}  \{ \bm{x}|\Phi,\bm{x}_{\rm pri},v_{x_{\rm pri}}\!\},\\ \nonumber \\
 &{\rm Orth:}& v_{x_{\rm orth}} &= {\cal O}_{\rm SE}(v_{x_{\rm post}},v_{x_{\rm pri}}),\\
 && \bf{x}_{\rm orth} &={\cal O}(\bm{x}_{\rm post},v_{x_{\rm post}};\bm{x}_{\rm pri},v_{x_{\rm pri}}),\\\nonumber
\\
&{\rm DCT:} &  \bm{z}_{\rm pri} &= \bm{F} \bm{x}_{\rm orth},\quad
  v_{z_{\rm pri}}  = v_{x_{\rm orth}}, \label{equ.OAMP.FFT}  
\end{align}
\ES
where orthogonalization  is defined as 
\BS\label{Eqn:orth}
\begin{align}
    v_{\rm orth} &= {\cal O}_{\rm SE}(v_{\rm post},v_{\rm pri}) \nonumber\\
    &\equiv \left(v_{\rm post}^{-1}-v_{\rm pri}^{-1}\right)^{-1},\\
     \bm{u}_{\rm orth}&={\cal O}(\bm{u}_{\rm post},v_{\rm post};\bm{u}_{\rm pri},v_{\rm pri})\nonumber\\
    &   \equiv v_{\rm orth}\left(v_{\rm post}^{-1}\bm{u}_{\rm post}-v_{\rm pri}^{-1}\bm{u}_{\rm pri}\right),
\end{align}
\ES 
and the symbol-by-symbol de-clipping and the section-by-section de-modulation are calculated by  
\BS\label{1}
\begin{align}
&\gamma_\epsilon(z_{\rm pri})=\!\!\int\!\! zP(z|\Gamma,z_{\rm pri},v_{z_{\rm{pri}}})dz,\\
&\gamma_\epsilon^{\rm SE}(v_{z_{\rm{pri}}})=\!\!\int\!\! z^2P(z|\Gamma,z_{\rm pri},v_{z_{\rm{pri}}})dz-\gamma^2(z_{\rm pri}),\\
&\phi(\bm{x}_{\rm s}^{\rm pri})=\!\!\int\!\! \bm{x}_{\rm s}P(\bm{x}_{\rm s}|\Phi,\bm{x}_{\rm s}^{\rm pri},v_{x_{\rm{pri}}})d\bm{x}_{\rm s},\\
&\phi^{\rm SE}(\!v_{x_{\rm pri}}\!)\!=\!\tfrac{1}{B}\!\!\!\int\!\! \left\|\bm{x}_{\rm s}\!\!-\!\phi(\bm{x}_{\rm s}^{\rm pri})\right\|^2\!\!P(\bm{x}_{\rm s}|\Phi\!,\bm{x}_{\rm s}^{\rm pri},v_{x_{\rm{pri}}})d\bm{x}_{\rm s}.
\end{align}
\ES
The subscript ``$\rm s$'' denotes a section of the corresponding variance vector (see \eqref{Eqn:block_sparse}).}


\underline{\textbf{State Evolution (SE):}} 
 SE has been strictly proven to be a numerical tool \cite{Rangan2016, Takeuchi2017} that can asymptotically characterize the performance of OAMP decoding with any unitarily-invariant matrices. 
 
  {For simplicity, we let $v_z=v_{z_{\rm pri}}=v_{x_{\rm orth}}$ and $v_x=v_{x_{\rm pri}}=v_{z_{\rm orth}}$. The transfer curves of the orthogonal de-clipping and orthogonal demodulation are given by
  \BS\label{Eqn:SE}
 \begin{align}
 \!\!\!& v_x \!= \!\gamma_{\rm orth}^{\rm SE}(v_z,\epsilon, \lambda)\!\equiv \! {\cal O}_{\rm SE}\!\left(\delta\gamma_\epsilon^{\rm SE}(v_z)+(1\!-\!\delta)v_z,v_z\right)\\
 \!\!\!& \hspace{2.6cm}=\!v_z\! \big[ \delta^{-1}\big(1\!-  \! \lambda \gamma_{\epsilon}^{\rm SE}(v_z)/ v_z\big)^{\!-1}\!\!\! -1 \big],\label{equ.SE.declip}\\
 \!\!\!&  v_z = \phi_{\rm orth}^{\rm SE}(v_x)\equiv {\cal O}_{\rm SE}\left(\phi^{\rm SE}(v_x),v_x\right).\label{equ.SE.demod}
 \end{align}
 \ES 
 where ${\cal O}_{\rm SE}(\cdot)$ is given in \eqref{Eqn:orth}. In practice, the local transfer curves $\gamma_{\rm orth}^{\rm SE}(\cdot)$ and $\phi_{\rm orth}^{\rm SE}(\cdot)$ can be obtained by local Monte Carlo simulations, using the IID Gaussian property of OAMP.
}

\section{ {Irregularly clipped SR Code}}
  {In this section, we propose an irregular clipping technique to further improve the SER performance of the clipped SR codes. Using SE, we optimize the distribution of irregular clipping thresholds to minimize the SER of SR codes.}

\subsection{ {Irregular Clipping}}\label{Gamma'}
 {In the regular clipping in Section \ref{Sec:RC_SRC}, all the elements are clipped based on the same threshold. Differently, in irregular clipping, different clipping thresholds can be applied to different symbols, which provides more optimization space for the SR codes and thus results in better performance.} 

 {Define the irregular clipping function as
\BE\label{Eqn:Irr_clip}
 \bm{c}={\rm clip}^{\rm Irr}_{\bm{\epsilon
},\bf{\lambda}}(\bm{z}).
\EE 
where $\bf{\epsilon}=[\epsilon_1 \cdots\epsilon_K]$. Specifically,  $\bm{z}$ is partitioned into $K$ subvectors $\{\bm{z}_1 \cdots\bm{z}_K\}$, where  $\bm{z}_k$ is clipped with threshold $\epsilon_k$ and has length $\lambda_kM$. $\bf{\lambda}=[\lambda_1 \cdots\lambda_K]$ denotes the threshold distribution of $\bf{\epsilon}$ with $0\leq\lambda_k\leq1$ and $\sum_{k=1}^K\lambda_k=1$. Therefore, the codeword length is still $M$.  Similarly, $\bm{c}$ is partitioned into $\{\bm{c}_1 \cdots\bm{c}_K\}$. Then, the irregular clipping function in \eqref{Eqn:Irr_clip} is given by
\BE\label{equ.irr_clipping}
\bm{c}_k = \alpha_k \, {\rm clip}_{\epsilon_k}(\bm{z}_k),\quad k=1,\dots,K,
\EE
where ${\rm clip}_{\epsilon_k}(\cdot)$ is defined in \eqref{Eqn:clip} and $\alpha_k$ is a power compensation parameter for clipping function ${\rm clip}_{\epsilon_k}(\cdot)$. Similarly to \eqref{equ.clipped_SR}, we define an irregularly clipped SR code as
\begin{subequations}\label{eqn:Ic_SR}
\begin{align}
&\tilde{\Gamma}:\quad\bm{y}={\rm clip}^{\rm Irr}_{\bm{\epsilon
},\bf{\lambda}}(\bm{z})+\bm{n},  \vspace{1.5mm}\label{eqn:Ic_SR_a}\\ 
& \Psi:\quad\bm{z} = \bm{Fx}, \label{eqn:Ic_SR_b}
\\
&\Phi:\quad
\bf{x}_l \sim P(\bm{x}_l),\;\; \forall l, \label{eqn:Ic_SR_c}
\end{align}
\ES
where ${\rm clip}^{\rm Irr}_{\bm{\epsilon
},\bf{\lambda}}(\cdot)$ is given in \eqref{equ.irr_clipping}, and \eqref{eqn:Ic_SR_b} and \eqref{eqn:Ic_SR_c} are the same as those in \eqref{equ.clipped_SR}.} 

\subsection{ {OAMP Decoding for Irregularly Clipped SR codes}}
 Let $\bm{z}_{\rm post}=[\bm{z}^{\rm post}_1 \cdots, \bm{z}^{\rm post}_K]$, $\bm{z}_{\rm pri}=[\bm{z}^{\rm pri}_1 \cdots, \bm{z}^{\rm pri}_K]$ and $\bm{z}_{\rm orth}=[\bm{z}^{\rm orth}_1 \cdots, \bm{z}^{\rm orth}_K]$. In the decoding of irregularly clipped SR codes, the  de-clipping is performed for each subvector. Thus, the de-clipping in \eqref{equ.OAMP} is replaced by: $ \forall k$,
 \BS
 \begin{align} 
 &\!\!\! {\rm Irr\!-\!declip}: &\!\!\! \bm{z}_k^{\rm post}  &= {\gamma}_{\epsilon_k}(\bm{z}_k^{\rm pri}), \label{Eqn:irr_declip}\\
    &\!\!\!& \!\!\!\!\!v_{z_{\rm post}}  &=  \sum\limits_{k=1}^{K}  \lambda_k {\gamma}_{\epsilon_k}^{\rm SE}(v_{z_{\rm pri}}), \label{Eqn:irr_declip_v}   
 \end{align} \ES
where $\gamma_{\epsilon_k}(\cdot)$ and $\gamma_{\epsilon_k}^{\rm SE}(\cdot)$ are respectively given in \eqref{equ.OAMP.Declip} and \eqref{equ.OAMP.Declip_v}. The other parts (e.g. demodulation, orthogonalization and DCT/IDCT) of the OAMP decoding of irregularly clipped SR codes are the same as those in \eqref{equ.OAMP}.

\underline{\textbf{State Evolution (SE):}} In irregularly clipped SR codes, the transfer curve of orthogonal irregular de-clipping is given by
\BS\label{Eqn:SE_irr} \begin{align}
 v_x &= \tilde{\gamma}_{\rm orth}^{\rm SE}(v_z,\bf{\epsilon},\bf{\lambda})\\
 &\equiv  \Big[\Big(\delta \!  \sum\limits_{k=1}^{K} \! \lambda_k  \gamma_{\epsilon_k}^{\rm SE}(v_z) + (1-\delta) v_z\Big)^{-1} - v_z^{-1}\Big]^{\!-1} \\
 &= v_z \Big[ \delta^{-1}\Big(1-\sum\limits_{k=1}^{K} \! \lambda_k  \gamma_{\epsilon_k}^{\rm SE}(v_z)/ v_z\Big)^{-1} -1 \Big],
 \end{align}
\ES
where $\gamma_{\rm orth}^{\rm SE}(v_z,\epsilon)$ is given in \eqref{equ.SE.declip}. The transfer curve of orthogonal demodulation is the same as the regularly clipped SR codes in \eqref{equ.SE.demod}.   

\underline{\textbf{Complexity:}} It is easy to see that the computational complexity of irregularly clipped SR codes is almost the same as that of regularly clipped SR codes, dominated by DCT/IDCT with complexity $\mathcal{O}(N\log N)$ per iteration.

\subsection{ {Optimization of Irregular Clipping Thresholds}}\label{opt} 
 {In this subsection, given the irregular clipping  threshold vector $\bf{\epsilon}$, we will discuss the optimization $\bf{\lambda}=[\lambda_1 \cdots \lambda_K]$. Intuitively, the distribution vector $\bf{\lambda}$ controls the contribution of the different clipping thresholds. Therefore, we can optimize  $\bf{\lambda}$  to minimize the SER performance of irregularly clipped SR codes. Equivalently, given ${\rm SNR}$, we can maximize the minimal gap between the transfer curves $\tilde{\gamma}_{\rm orth}^{\rm SE}$ and $\phi_{\rm orth}^{\rm SE}$ in $[v_{\min}, 1]$, where $v_{\min}$ is a small enough positive number to obtain the desired SER performance. Hence, the optimization problem can be described as
\BS\begin{align}
  {\mathcal P}_1:  \;\;  \mathop{\rm{max}}_{\bm{ \lambda}}  \mathop{\rm{min}}_{v\in[v_{\min},1]} \;\;&  {\phi_{\rm orth}^{{\rm SE}^{-1}}}\!(v)-  \tilde{\gamma}_{\rm orth}^{\rm SE}(v,\bf{\epsilon},\bm{\lambda}),\\
   {\rm s.t.}\;& \; \sum\limits_{k=1}^{K} \lambda_k = 1,\\
& \; 0\leq  \lambda_k\leq 1,
\end{align}\ES
where $\tilde{\gamma}_{\rm orth}^{\rm SE}(v,\bf{\epsilon},\bm{\lambda})$ is given in \eqref{Eqn:SE_irr}, and ${\phi_{\rm orth}^{{\rm SE}^{-1}}}(\cdot)$ is an inverse function of ${\phi_{\rm orth}^{{\rm SE}}}(\cdot)$ in \eqref{equ.SE.demod}. } 

 Notice that the objective function in ${\mathcal P}_1$ is a complicated function of $v$. Given $\bf{\lambda}$, it is very difficult to find the analytical solution of ${\mathcal P}_1$ on the continuous region ${v\in[v_{\min},1]}$. To overcome this problem, we consider minimization on the uniformly sampling points $\mathcal{V}=\{v_i\}$ on $[v_{\min},1]$ in log domain. In general, we set $v_{\min}=10^{-6}$ and the number of sampling points as 100, i.e., $|\mathcal{V}|=100$. Therefore, following \eqref{Eqn:SE_irr}, ${\mathcal P}_1$ can be rewritten to the following optimization problem:
\BS\begin{align}
 \!\! {\mathcal P}_2: \;\; \mathop{\rm{max}}_{\bm{ \lambda}} \; \mathop{\rm{min}}_{v_i\in{\mathcal{V}}} &\quad {\phi_{\rm orth}^{{\rm SE}^{-1}}}(v_i) - \tilde{\gamma}_{\rm orth}^{\rm SE}(v_i,\bf{\epsilon},\bm{\lambda}),\\
   {\rm s.t.}& \quad \sum\limits_{k=1}^{K} \lambda_k = 1,\\
& \quad 0\leq  \lambda_k\leq 1.
\end{align}\ES
Note that $\delta, v_i, \gamma_{\epsilon_k}^{\rm SE}(v_i)$ are all positive. It is easy to verify that $\{\tilde{\gamma}_{\rm orth}^{\rm SE}(v_i,\bf{\epsilon},\bm{\lambda}), \forall i\}$ (see \eqref{Eqn:SE_irr}) are concave functions of $\bf{\lambda}$. The minimization of concave functions is also concave \cite{Boyd2004}. In addition, the constraints in ${\mathcal P}_2$ are linear. Therefore, ${\mathcal P}_2$ is a concave problem of $\bf{\lambda}$ and thus it can be solved using standard convex optimization tools.  

\subsection{ {Transfer Chart Comparison of Regularly Clipped and Irregularly Clipped SR codes}}\label{SE_ana}
 \begin{figure}[t]
    \centering
     \includegraphics[width = 8.8cm]{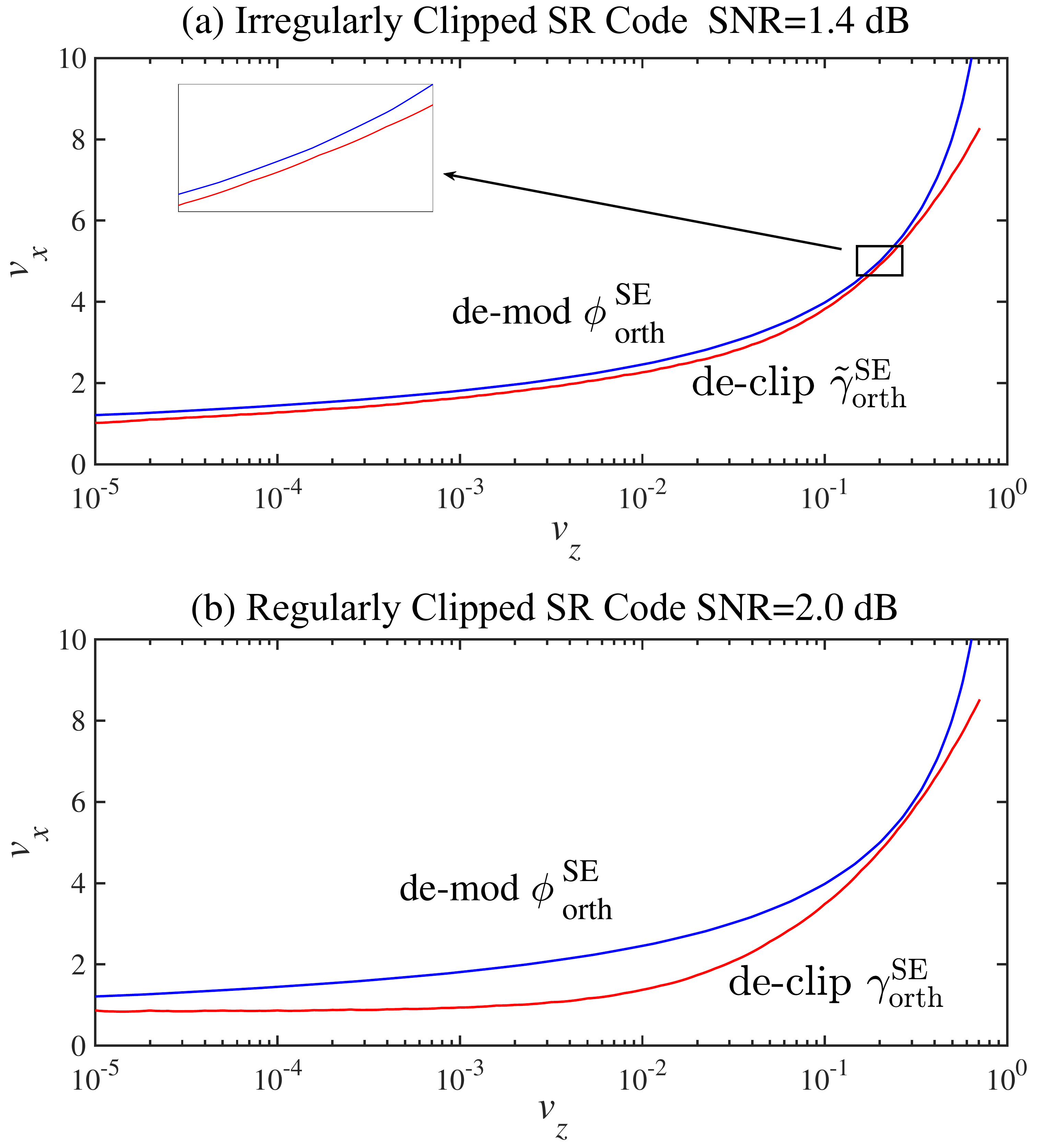} 
    \caption{ {SE transfer charts of (a) the optimized regularly clipped SR code with ${\rm SNR}=2$ dB and (b) the optimized irregularly clipped SR codes with ${\rm SNR}=1.3$ dB.  $B=64$, $L=2048$ and $R=0.5$. The optimal CR of regular clipping for regularly clipped SR codes is $-13$ dB.}}
    \label{Fig.SE}
\end{figure}
 {Fig. \ref{Fig.SE} compares the SE transfer charts of the optimized regularly clipped SR codes and the optimized irregularly clipped SR codes. As illustrated in Fig. \ref{Fig.SE}, to ensure the OAMP decoding converges to the target SER, the key is to create a decoding tunnel between the curves of demodulation and de-clipping. Let $v_{\min}$ be the target performance. There should be no fixed point in the region $v\leq v_{\min}$, since otherwise the
tunnel will be closed at $v > v_{\min}$. It is shown that the tunnel is opened at ${\rm SNR}=2$ dB for regularly clipped SR codes (see Fig. \ref{Fig.SE}(a)), and at ${\rm SNR}=1.3$ dB for irregularly clipped SR codes (see Fig. \ref{Fig.SE}(b)). In other words, following the SE transfer chart analysis, irregularly clipped SR codes achieve about $0.7$ dB gain in ${\rm SNR}$ comparing with regularly clipped SR codes for the target performance $v_{\min}=10^{-5}$.}

\section{Simulation Results}
 {Fig. \ref{Fig.simulation_result} gives the SERs of non-clipped SR code, optimized regularly clipped SR code and optimized irregularly clipped SR code. The parameters are set as $B=64$, $L=2048$, $R = 0.5$ (code rate), $M = 24,576$ (code length) and ${\rm SNR}=\sigma^{-2}_n$. The optimal CR of regularly clipped SR codes is $-13$ dB. The irregularly clipped SR codes involve 19 candidate CRs for irregular clipping. For the irregularly clipped SR codes in Fig. \ref{Fig.simulation_result}, the CRs and corresponding distribution vector $\lambda$ are optimized for each simulated SNR (see the concave optimization in Section \ref{opt}). The maximum number of iterations is 120. As we can see in Fig.~\ref{Fig.simulation_result}, the proposed irregularly clipped SR codes achieve better SER performance than regularly clipped and non-clipped. Comparing with regularly clipped SR codes \cite{Liang_cl2017},  irregularly clipped SR codes achieve about 0.4 dB gain in ${\rm SNR}$ at SER=$10^{-5}$.}

 \begin{figure}[t]
    \centering
    \includegraphics[width =8.8cm]{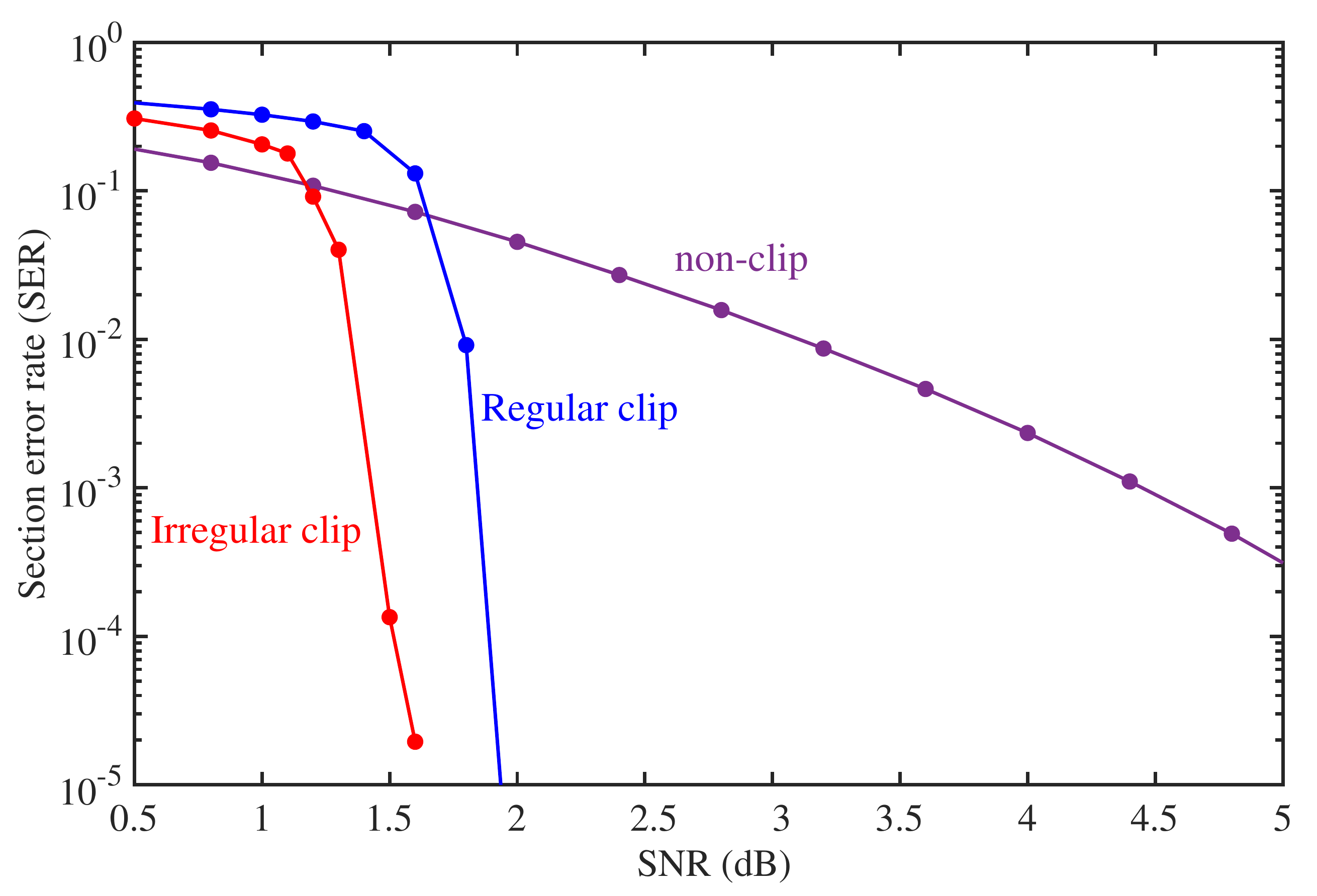}
    \caption{ {SER comparison between the non-clipped SR codes, optimized regularly clipped SR codes and optimized irregularly clipped SR codes. $B=64$, $L=2048$, $R = 0.5$ (code rate), $M = 24,576$ (code length) and ${\rm SNR}=\sigma^{-2}_n$. The CRs and its distribution $\bm{\lambda}$ are given in Table \ref{tab.lambda}. The maximum number of iterations is 120. The optimal CR of regular clipping for regularly clipped SR codes is $-13$ dB.}}
    \label{Fig.simulation_result}
\end{figure}
 Fig.~\ref{Fig.rates} gives the SERs of optimized irregularly clipped SR codes with code rates $R=\{0.2, 0.4, 0.6,0.8, 1\}$. The parameters are set as  $B=64$, $L=2048$, and $M = \{61440, 30720, 20480, 15360, 12288\} $ (code lengths). The maximum number of iterations is 100. As can be seen, the optimized irregularly clipped SR codes exhibit characteristic waterfall behavior for a range of rate from 0.2 (low rate) to 1 (high rate). Every point in Fig.~\ref{Fig.rates} uses  different CRs and $\bm{\lambda}$. Table~\ref{tab.lambda} shows the $\bm{\lambda}$ of each code rate for the points whose SER is close to $10^{-5}$.  

 \begin{figure}[t]
    \centering
    \includegraphics[width =8.75cm]{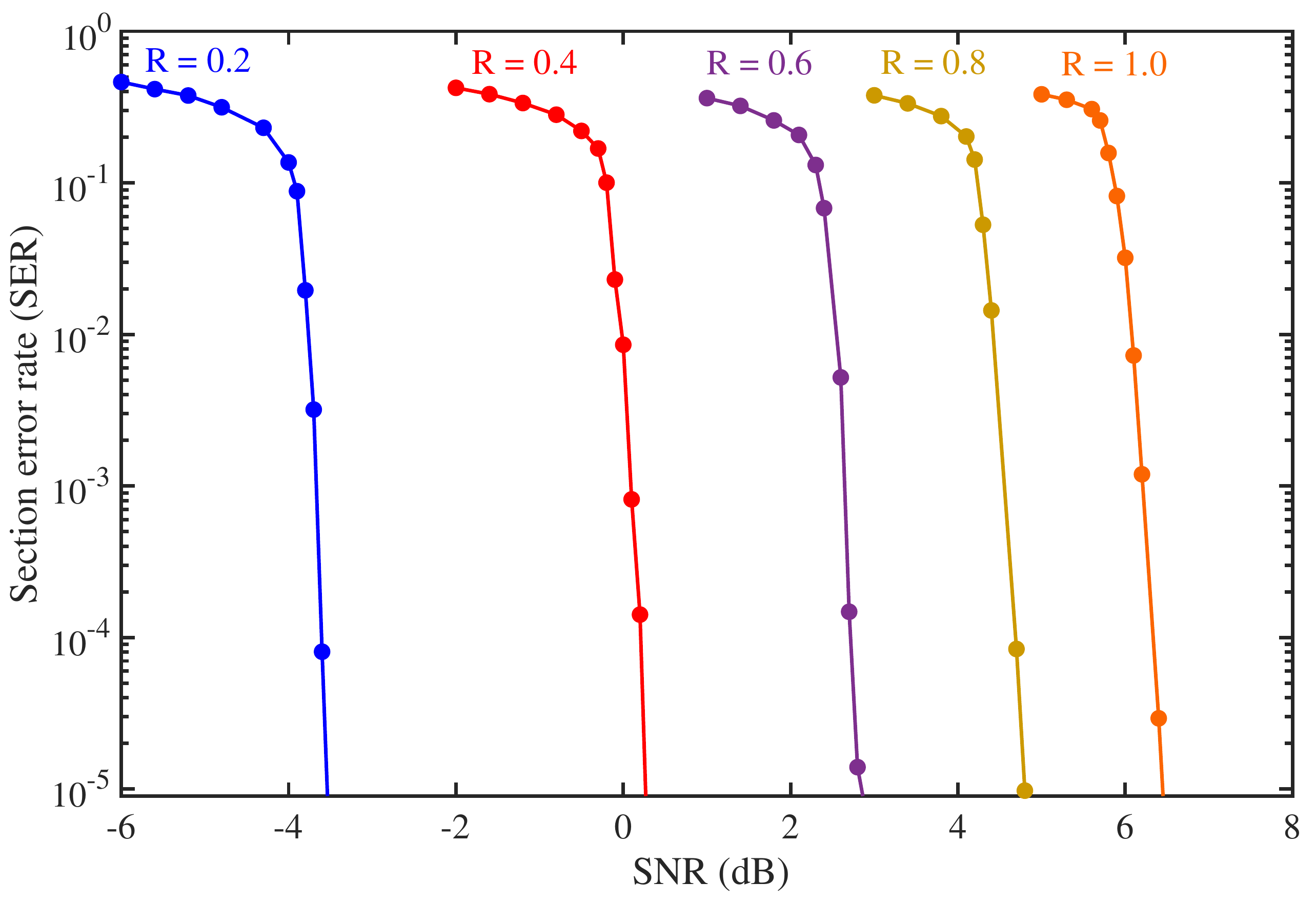}
    \caption{ SERs of the optimized irregularly clipped SR codes with different code rates. $B=64$, $L=2048$,  $R=\{0.2, 0.4, 0.6,0.8, 1\}$ (code rates), $M = \{61440, 30720, 20480, 15360, 12288\} $ (code lengths) and ${\rm SNR}=\sigma^{-2}_n$. CRs and the distribution are given in Table~\ref{tab.lambda}. The maximum number of iterations is 100.}
    \label{Fig.rates}
\end{figure}
\begin{table}[]
\centering
\caption{optimized coefficient CRs and $\bm{\lambda}$ }\footnotesize
\begin{tabular}{ccc}
    \hline R &SNR (dB)&CR (dB) and $\bm{\lambda}$  \\ \hline
    \multirow{2}*{0.2} & \multirow{2}*{-3.6}  &\!\!-300 \qquad -12 \qquad\;-10 \qquad\;-8\\ &&0.04460\; 0.27394\; 0.38313\; 0.29831\\
    \hline   \multirow{4}*{0.4} & \multirow{4}*{0.2} &\!\!-300 \quad\;\;\;-30 \qquad\;\;-16 \qquad\;-10 \qquad\;\;-8 \ \\
    &&0.02590\; 0.05297\; 0.00756\; 0.00145\; 0.17107\\
    &&\!\!-6 \ \ \qquad -2 \qquad\ \;0 \qquad\ \  \;4\\&&0.51890\; 0.13035\; 0.00790\; 0.08390\\
    \hline\multirow{4}*{0.5} & \multirow{4}*{1.6} &-300 \qquad\;\;-24\qquad\;\;-22\qquad\;\;-18\qquad\;\;-6\\
    &&0.01251\; 0.12000\; 0.00027\; 0.00293\; 0.00031\\
    &&-5\qquad\;\;-4\qquad\;\;-3\qquad\;\;-2\\
    &&0.07169\; 0.56832\; 0.16883\; 0.05508\\
      \hline   \multirow{4}*{0.6} & \multirow{4}*{2.8}&-300\  \ \qquad-30\ \  \qquad-19\ \ \qquad-16\\&&0.02048\; 0.01288\; 0.13391\; 0.00462\\&&-12\ \   \qquad-4 \ \ \ \qquad-2\ \ \ \qquad0\\&&0.00034\; 0.19962\; 0.26163\; 0.36647\\
       \hline   \multirow{4}*{0.8} & \multirow{4}*{4.8}&-300   \qquad-30\   \qquad     -22\ \ \qquad-16  \ \qquad-14 \ \ \\&&0.01207\; 0.02881\; 0.01619\; 0.07276 \ 0.04741\\&&-10 \quad \qquad0 \quad     \qquad2   \quad  \qquad6 \quad    \qquad8\\&& 0.07536\; 0.06970\; 0.43372\ 0.17212\; 0.07178\\
        \hline   \multirow{4}*{1.0} & \multirow{4}*{6.4}&-300\ \qquad-30\ \ \qquad-22\  \qquad-20 \ \ \\&&0.00899\; 0.03115\; 0.00010\; 0.00380\\&& -16\ \ \qquad-14\ \ \qquad-6\ \qquad300\ \ \\&&0.00037\; 0.16628\; 0.00015\; 0.78912 \\ \hline
    \end{tabular}
    \label{tab.lambda}
\end{table}
\section{Conclusion}
This paper investigated irregular clipping for SR codes with OAMP decoding. Using SE analysis, we constructed a convex optimization problem to minimize the SER of irregularly clipped SR codes by optimizing the distribution of the clipping thresholds. As a result, the proposed irregularly clipped SR codes achieve a better SER performance than the existing regularly clipped SR codes. When compared with regularly clipped SR codes, irregularly clipped SR codes achieve about 0.4 dB gain in SNR at code length $\approx 2.5\times 10^4$ and ${\rm SER}\approx 10^{-5}$. Furthermore, irregularly clipped SR codes are robust in a wide range of code rates, e.g., from 0.2 (low rate) to 1 (high rate). 
 
\section*{Acknowledgment}  
The authors thank Shansuo Liang for the discussions that have improved the quality of this work greatly.

\end{document}